\documentclass[aip,reprint]{revtex4-1}
\usepackage{graphicx}

\draft % marks overfull lines with a black rule on the right

\begin{document}

%\preprint{}

\title{Dielectric properties of relaxor-ferroelectric ceramic and single crystal Pb(In$_{\text{1/2}}$Nb$_{\text{1/2}}$)O$_{\text{3}}$-Pb(Mg$_{\text{1/3}}$Nb$_{\text{2/3}}$)O$_{\text{3}}$-PbTiO$_{\text{3}}$ at cryogenic temperatures} %Title of paper



\author{P.M. Shepley}
%\email[]{p.m.shepley@leeds.ac.uk}
%\homepage[]{https://engineering.leeds.ac.uk/staff/1002/Dr_Philippa_Shepley}
%\thanks{}
%\altaffiliation{School of Physics and Astronomy, University of Leeds, Leeds, U.K.}
\affiliation{School of Chemical and Process Engineering, University of Leeds, Leeds, U.K.}

\author{L.A.Stoica}
%\email[]{l.a.stoica@leeds.ac.uk}
%\homepage[]{}
%\thanks{}
%\altaffiliation{Thales UK, Templecombe, U.K.}
\affiliation{School of Chemical and Process Engineering, University of Leeds, Leeds, U.K.}

\author{Y. Li}
%\email[]{pmylia@leeds.ac.uk}
%\homepage[]{}
%\thanks{}
\affiliation{School of Chemical and Process Engineering, University of Leeds, Leeds, U.K.}

\author{G. Burnell}
%\email[]{G.Burnell@leeds.ac.uk}
%\homepage[]{http://www.stoner.leeds.ac.uk/people/gb}
%\thanks{}
\affiliation{School of Physics and Astronomy, University of Leeds, Leeds, U.K.}

\author{A.J. Bell}
%\email[]{A.J.Bell@leeds.ac.uk}
%\homepage[]{https://engineering.leeds.ac.uk/staff/238/Professor_Andrew_Bell}
%\thanks{}
\affiliation{School of Chemical and Process Engineering, University of Leeds, Leeds, U.K.}


\date{\today}

\begin{abstract}
We investigate the low temperature behaviour of Pb(In$_{\text{1/2}}$Nb$_{\text{1/2}}$)O$_{\text{3}}$-Pb(Mg$_{\text{1/3}}$Nb$_{\text{2/3}}$)O$_{\text{3}}$-PbTiO$_{\text{3}}$ using dielectric permittivity measurements. We compare single crystal plates measured in the [001] and [111] directions with a polycrystalline ceramic of the same composition. Poled crystals behave very differently to unpoled crystals, whereas the dielectric spectrum of the ceramic changes very little on poling. A large, frequency dependent dielectric relaxation seen in the poled [001] crystal around 100 K is much less prominent in the [111] crystal, and doesn’t occur in the ceramic. Preparation conditions and the microstructure of the material play a role in the low temperature dynamics of relaxor-ferroelectric crystals.
\end{abstract}


\maketitle %\maketitle must follow title, authors, abstract and \pacs

\section{Introduction}
%\label{}

Single crystal relaxor-PbTiO$_{\text{3}}$ ferroelectric materials can have exceptionally high piezoelectric properties at room temperature. Their large piezoelectric and dielectric constants, along with low dielectric losses are desirable for a wide range of applications \cite{Park1997,Zhang2012,Zhang2015}. Much of the recent effort to understand the origins of the excellent room temeprature properties of relaxor-PbTiO$_{\text{3}}$ materials has focused on understanding the piezo- or dielectric behaviour of the materials below room temperature. A relaxation step feature in piezoelectric and dielectric properties at low temperatures have been reported in single crystal relaxor-PbTiO$_{\text{3}}$ samples, \cite{Martin2012,Bukhari2014,Li2016} in addition to the characteristic relaxor-ferroelectric dielectric peaks above room temperature \cite{Wang2009,Wang2014,Li2016,Takenaka2017}. 

For rhombohedral Pb(Mg$_{\text{1/3}}$Nb$_{\text{2/3}}$)O$_{\text{3}}$-PbTiO$_{\text{3}}$ (PMN-PT) crystals, Martin et al and Li et al  \cite{Martin2012,Porokhonskyy2010} have shown that at around 200 K the reduction in dielectric permittivity and piezoelectricity with temperature levels off, before dropping sharply between 100 K and 20 K. The drop in permittivity is associated with a peak in the dielectric loss, both of which show a large variation as a function of driving frequency. The step feature suggests a "freezing out" of temperature activated dynamics, which has been cited as evidence that the persistence of polar nano-regions down to lower temperatures gives relaxor-PbTiO$_{\text{3}}$ materials their high room temperature properties\cite{Li2016,Li2017}. 

Low temperature dielectric data from different ferroelectric and relaxor-ferroelectric materials show a wide range of anomalies and features at cryogenic temperatures. The large dielectric relaxation feature highlighted by Li et al \cite{Li2016} is not always present in relaxor-PbTiO$_{\text{3}}$ single crystals \cite{Lente2004,Wang2009,Hentati2015}. Studies on PMN-PT suggest that the material composition and poling state \cite{Wang2009} influence the size, shape and presence of a low temperature feature. Work on PMN-PT ceramics \cite{Singh2007,Li2008} showed broad dielectric loss anomalies that appear more similar to some data on lead zirconate titanate (PZT) based ceramics than relaxor-PbTiO$_{\text{3}}$ single crystals \cite{Zhang1983,Zhang1994,Thiercelin2010}. In PZT ceramics, freezing out of the motion of domain walls is used to explain broad peaks in the dielectric loss spectra \cite{Li2014}. There are dielectric data on Fe doped PZT ceramics \cite{Arlt1987} showing a step feature with frequency dispersion very similar to that seen in PMN-PT \cite{Martin2012,Li2016}. Arlt et al found good agreement between this data and their domain wall dynamics model \cite{Arlt1987}.

The low temperature, rhombohedral phase of BaTiO$_{\text{3}}$ can give dielectric data that peak to anomalously high values, then reduce close to 0 K \cite{Akishige1998,Akishige2004,Wang2007,Wang2011}. The peaks are similar to those seen in PMN-PT in their shape and frequency dispersion. The presence of this low temperature relaxation behaviour in ferroelectric BaTiO$_{\text{3}}$ single crystals has been shown to vary between crystals grown by different methods, and depending on the crystals' electric field and temperature histories \cite{Akishige1998,Wang2007}. Wang et al have shown that images of different domain states can be linked to differences in the size of the dielectric constant peak in the rhombohedral phase of BaTiO$_{\text{3}}$ single crystals \cite{Wang2011}.

In order to understand the origins of low temperature anomalies and their relationship to room temperature properties, data on materials in a range of conditions are required. Here we investigate the relaxation step in relaxor-PbTiO$_{\text{3}}$ materials by considering polycrystalline ceramic and single crystal Pb(In$_{\text{1/2}}$Nb$_{\text{1/2}}$)O$_{\text{3}}$-Pb(Mg$_{\text{1/3}}$Nb$_{\text{2/3}}$)O$_{\text{3}}$-PbTiO$_{\text{3}}$ (PIN-PMN-PT). We investigate the effects of poling the material and find that the large relaxation step only becomes apparent when the single crystal is poled.

\begin{figure}[ht]
\centering
\includegraphics[width=7cm]{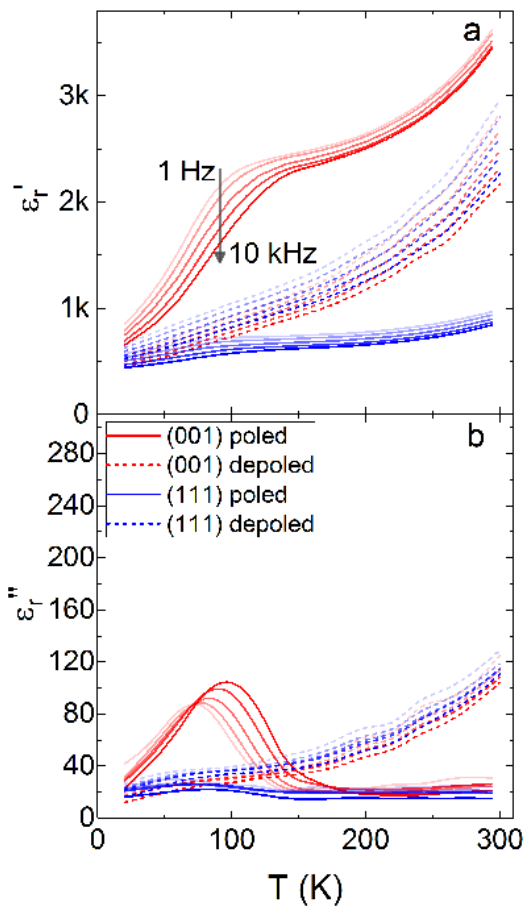}
\caption{The a) real and b) imaginary parts of the dielectric permittivity in (001) and (111) cut PIN-PMN-PT single crystals are shown at temperatures from 20 $K$ to 300 $K$. The solid, red lines are for the poled (001) cut and the dashed, red lines are for the depoled (001) cut. The solid, blue lines are for the poled (111) cut and the dashed, red lines for the depoled (111) cut.}
\label{fig:permTcrystal}
\end{figure}

\begin{figure}[ht]
\centering
\includegraphics[width=7cm]{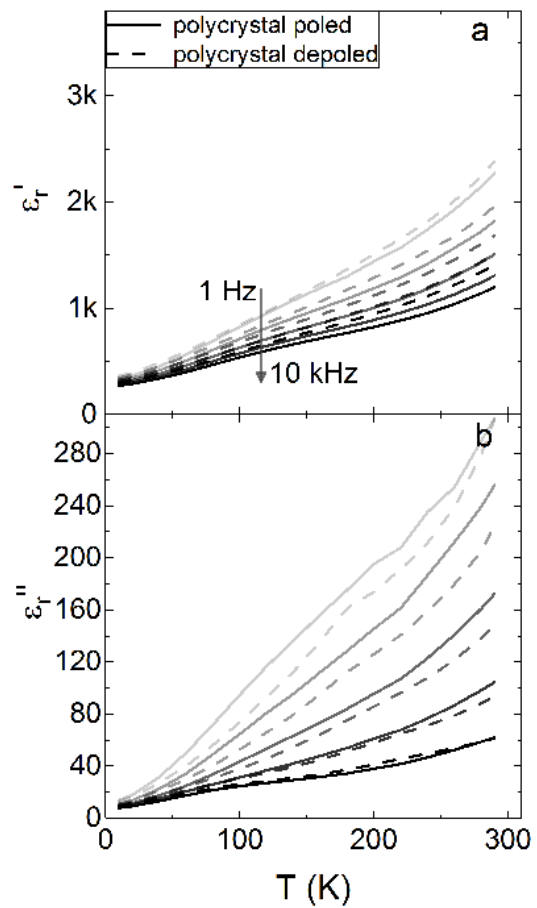}
\caption{The a) real and b) imaginary parts of the dielectric permittivity in a PIN-PMN-PT polycrystalline ceramic are shown at temperatures from 10 $K$ to 300 $K$. The solid lines are for the poled ceramic and the dashed lines are for the depoled ceramic.}
\label{fig:permTpoly}
\end{figure}

\section{Experimental Details}
%\label{}

We have studied relaxor-ferroelectric PIN-PMN-PT below room temperature. The data shown here are from crystal plates cut with (001) and (111) faces, and from a polycrystalline ceramic pellet. To make the samples, powdered material was prepared by mixed oxide methods. The polycrystalline ceramic pellet was formed by sintering and the crystal was grown by Bridgman technique \cite{Stoica2016}.   Silver epoxy was painted onto the crystal and pellet main faces and cured at 770 K to form electrodes. The (001) cut crystal was 1.117 mm thick and had an electrode area of 0.2019 cm$^{2}$, the (111) cut crystal was 1.200 mm thick and had an electrode area of 0.2601 cm$^{2}$, and the ceramic pellet was 1.720 mm thick and had an electrode area of 0.7557 cm$^{2}$. The nominal composition of the material, PIN$_{\text{0.28}}$-PMN$_{\text{0.40}}$-PT$_{\text{0.32}}$, was chosen to be close to the morphotropic phase boundary (MPB) and give a rhombohedral structure at room temperature.

The samples were all prepared for measurements in the poled state by first annealing to 850 K, a point well above any phase transitions or dielectric maxima, then allowing them to cool to room temperature. We poled the samples by heating them to 370 K, then applying an electric field of 1 kV/mm while the samples cooled to room temperature. The elevated temperature is high enough to lower the energy barrier for domain re-orientation, but is below any phase transitions. All samples showed piezoelectric resonance peaks at high frequency, indicating that they were properly poled. The measurements in a depoled state were taken after the samples had been annealed to a point (above 500 K) where they no longer showed a spontaneous polarisation or piezoelectric resonance peaks.

The real and imaginary parts of the dielectric permittivity, $\varepsilon _{\text{r}}'$ and $\varepsilon _{\text{r}}''$, were measured with a Solartron impedance analyser and XM-Studio MTS software. The crystals were mounted in an Oxford Microstat, where the temperature was swept at a rate of 2 K/minute between 10 K and 300 K. A driving voltage with an rms value of 2 V was applied at a range of frequencies between 10 kHz and 0.05 Hz, and the response was measured. The full set of data is available from (DOI to be inserted).

\section{Results}
%\label{}

The real and imaginary parts of the dielectric permittivity, $\varepsilon _{\text{r}}'$ and $\varepsilon _{\text{r}}''$, measured for the (001) and (111) cut PIN-PMN-PT crystals, in a poled and depoled state, are plotted in Figure \ref{fig:permTcrystal}. We see the same features in the dielectric properties of (001) poled rhombohedral PIN-PMN-PT as have been reported for PMN-PT \cite {Li2016}, but we find differences between the two crystal cuts and there are large differences between the depoled and poled crystals. 

The real part of the permittivity $\varepsilon _{\text{r}}'$ of the (001) cut crystal increases when the sample is poled, whereas for the (111) cut $\varepsilon _{\text{r}}'$ decreased after poling. For both crystals, the frequency dispersion at room temperature is reduced by poling, however the room temperature values of permittivity are very different. 

The poled (001) crystal has the permittivity step feature seen in PMN-PT, where the rate of decrease of $\varepsilon _{\text{r}}'$ as the sample is cooled slows at 200 K, then increases around 100 K, so that $\varepsilon _{\text{r}}'$ drops sharply. The feature is also present to some degree in the (111) crystal, although the size and sharpness of the drop is much less significant than in the (001) crystal. 

The imaginary part of the permittivity $\varepsilon _{\text{r}}''$ is low at room temperature in the poled crystals. There is very little change in $\varepsilon _{\text{r}}''$ from the room temperature value in the (111) crystal. In the (001) crystal the $\varepsilon _{\text{r}}'$ step feature is associated with a large peak in $\varepsilon _{\text{r}}''$. 

The behaviour with temperature of the two depoled PIN-PMN-PT crystals, (001) and (111) cut, is almost identical. There is a large variation in the relative permittivity $\varepsilon _{\text{r}}'$ of depoled PIN-PMN-PT with driving frequency. The frequency dispersion is largest at room temperature, then below 150 K the frequency dispersion begins to decrease. The imaginary part of the permittivity $\varepsilon _{\text{r}}''$ follows a similar function of temperature as $\varepsilon _{\text{r}}'$, dropping to approximately 20 $\%$ of its room temperature value by 20 K, with no prominent step features.

The polycrystal behaves in a similar way to the depoled crystals. The real and imaginary parts of the dielectric permittivity, $\varepsilon _{\text{r}}'$ and $\varepsilon _{\text{r}}''$, were measured for the ceramic in a poled and depoled state and the results are shown in Figure \ref{fig:permTpoly}. There is a substantial difference as a function of frequency in the dielectric data (both real and imaginary parts), which closes as the temperature approaches 0 K. The slopes of relative permittivity $\varepsilon _{\text{r}}'$ and the imaginary part of the permittivity $\varepsilon _{\text{r}}''$ are almost constant over the measured temperature range, although the data at 1 Hz are considerably steeper than the data at 10 kHz. In the range of temperature and frequency in Figure \ref{fig:permTpoly} there is very little change in the permittivity spectra of the polycrystal when it is poled compared to when it is depoled.

\begin{figure}
\includegraphics[width=7cm]{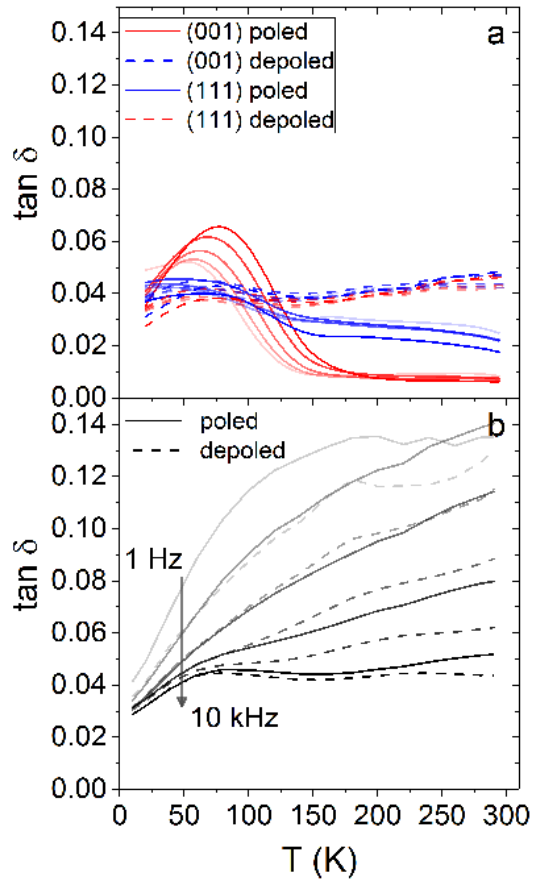}
\caption{The dielectric loss in a) (001) and (111) cut PIN-PMN-PT single crystals and b) a polycrystalline ceramic are shown at temperatures from 10 $K$ to 300 $K$. The solid lines are for the poled material and the dashed are for the depoled material. The red lines are for the (001) cut, the blue lines are for the (111) cut and the black lines are for the polycrystalline ceramic.}
\label{fig:tand}
\end{figure}

In addition to the real and imaginary parts of the dielectric permittivity, we also show the dielectric loss tangent for all samples in Figure \ref{fig:tand}. For the single crystal samples the low temperature features in the imaginary permittivity (Figure \ref{fig:permTcrystal}b) and the dielectric loss (Figure \ref{fig:tand}a) are qualitatively similar. The peak in the (001) data represents a maximum in the energy lost when changing the polarisation direction.

For the polycrystalline ceramic the dielectric loss in Figure \ref{fig:tand}b shows a more prominent low temperature effect than the permittivity in Figure \ref{fig:permTpoly}. The dielectric loss in the polycrystal changes less than $\varepsilon _{\text{r}}''$ close to room temperature, and we then see larger changes at lower temperatures. Loss data at frequencies below 1 kHz reduces more steeply below 150 K, whereas for higher frequency data there is small bump that could indicate a high frequency process that freezes out below 100 K.

\section{Discussion}
%\label{}

Low temperature dielectric relaxation data have been modelled and explained by freezing-out of dynamics associated with both domains \cite{Arlt1987,Wang2011} and polar nano-regions \cite{Li2017}. The preparation conditions of materials, including the crystal growth method, crystalline or polycrystalline nature and the field and temperature histories are all important factors in the low temperature dielectric response. 

Poling a crystal changes the domain configuration, aligning randomly oriented domains to give a macroscopic polarisation direction. Poling a PIN-PMN-PT crystal along [111] aligns the polarisation to an energetically favourable direction for the rhombohedral crystal structure, giving a single domain. During the permittivity measurement, we apply a small ac driving voltage along the same axis as the polarisation. The polarisation can only respond by changing in size, so the response is small, meaning we measure a lowered permittivity in the (111) poled crystal. 

The largest effect from low temperature relaxation is seen in the (001) crystal sample, where there is a domain state with a high degree of order. For a [001] poled rhombohedral PIN-PMN-PT crystal we expect four domain variants with polarization pointing along each of the $\{$111$\}$ axes, to the corners of the crystal's unit cell -- all with a component along [001]. In the (001) cut crystal the polarisation is predominantly not aligned to the same axis as the ac driving voltage. The polarisation domains can respond by rotating towards and away from the [001] direction, as well as by changing the magnitude of their polarisation, giving an increased permittivity in the (001) cut. 

The motion of domain walls can contribute to polarisation changes, and therefore to the permittivity. Both of the depoled crystals have eight rhombohedral domain variants. The depoled polycrystalline material has domains with polarisation pointing in all directions because of the range of orientations of crystal grains. Poling the polycrystal gives domains in all the directions that have some positive component along [001]. 

Poling PIN-PMN-PT \cite{Wang2014} and PMN-PT \cite{Kiat2002,Cao2006a} crystals with compositions close to the MPB has been shown to change the crystallographic symmetry. X-ray diffraction shows that field cooling a crystal that is initially rhombohedral at room temperature gives a monoclinic symmetry. A difference in crystal symmetry between the poled and depoled samples would change the availible domain states and polarisation orientations. However, there is also evidence that the apparent monoclinic structure in materials close to the MPB is due to the averaging of variations in local structure, such as a combination of rhombohedral and tetragonal nano-domains\cite{Shvartsman2004,Schonau2007,Kim2012}. Poling relaxor-PbTiO$_{\text{3}}$ materials close to the MPB may enhance nanoscale structural variations that emerge from compositional variations, giving rise to phase domains with different crystal symmetries and polarisation orientations.

Whether the polarisation-change mechanism involves reorientation within a domain or motion of domain walls depends on the energies needed to activate the processes. If all domains have an equal polarisation component along the direction of an applied electric field, there should be no difference in the activation energy to rotate the polarisation within domains, and there will be no energetic benefit to domain wall motion. If there are domains with differences in polarisation component along an applied electric field direction -- for example if there is a combination of rhomboherdral and tetragonal phase domains -- domain walls may move to expand domains whose polarisation is better aligned to the applied field.

The range of orientations in the unpoled and the polycrystalline samples make domain wall motion likely as a mechanism for polarisation change \cite{Arlt1987}. In these samples, domain wall contributions are a good candidate for deviations from the permittivity expected from Landau theory \cite{Wang2011}. The broad frequency dispersion in the polycrystalline material is likely to be a consequence of many types of domain walls with a large range activation energies that respond to the applied electric field at different frequencies.

The relaxation step that we measure in the PIN-PMN-PT (001) single crystal is very similar to both the step seen in PMN-PT \cite{Li2016,Li2017} (which has been modelled by dynamics of polar nano-regions) and in rhombohedral BaTiO$_{\text{3}}$ \cite{Wang2007,Wang2011} (which has been modelled by dynamics of domain walls). Comparing the permittivity versus temperature of single crystals and a polycrystal in poled and depoled states plotted in Figure \ref{fig:permTcrystal} and \ref{fig:permTpoly} shows that the low temperature step feature is only present in poled single crystals. The difference in the low temperature dielectric properties between single crystal and polycrystalline material with the same composition shows that the relaxation features reported here and by other researchers  may not be entirely a consequence of the relaxor-like behaviour of polar nano-regions in a ferroelectric matrix. 

It is not clear that the motion of ferroelectric domain walls can be used to explain the effects in the (001) cut crystal, since there are domain wall motion mechanisms for the polarisation in the polycrystal and in the depoled single crystals, which have very different low temperature dielectric features. If the relaxation step in PIN-PMN-PT were due to the freezing of domain wall dynamics, we might expect there to be no effect at all in the single domain (111) crystal. Instead we see a small relaxation step, although it's possible that this could indicate dynamics from domains at the crystal's surfaces, rather than in the bulk, or the existance of a small number of phase domains enhanced by poling.

The data we present here, along with that of other studies on single crystals and ceramics, suggest a mechanism that is dependent on preparation conditions and sample history. The key requirement is a dynamic process linked to polarisation. So far, mechanisms have been proposed that depend on the motion of domain walls \cite{Arlt1987,Wang2011} or polarisation changes of polar nano-regions \cite{Li2016,Li2017}. Since we have shown that these mechanisms don't account for all of our data, we suggest other candidates: the fluctuations of nanoscale ferroelectric domains, or the motion of phase domain walls that lie at the interface between nano-regions with different crystal symmetry and polarisation orientation.

\section{Conclusions}
%\label{}

We have presented dielectric data below room temperature for the relaxor-ferroelectric material PIN-PMN-PT. We compare PIN-PMN-PT measured in six different conditions: poled and depoled single crystal (001) cut, poled and depoled single crystal (111) cut, and poled and depoled polycrystalline ceramic. 

The large dielectric relaxation feature reported in relaxor-PbTiO$_{\text{3}}$ is only present in the poled single crystals, and is much more prominent in the multi-domain (001) cut than the single-domain (111). The differences between sample material under different conditions show that low temperature relaxations in relaxor-PbTiO$_{\text{3}}$ materials cannot be fully explained by a model based on the dynamics of polar nano-regions in a ferroelectric matrix \cite{Li2016,Li2017} or by a model based on the motion of domain walls \cite{Arlt1987}. The former model would suggest that the relaxation should be present in all the PIN-PMN-PT samples, and the latter would suggest that the relaxation should be present in all the unpoled samples, but not in the single domain (111) crystal. 

In addition to polarisation mechanisms for low temperature dynamics involving domain walls or polar nano-regions, mechanisms involving fluctuations of nanoscale ferroelectric domains and motions of phase domain walls could contribute low temperature dielectric features.

\begin{acknowledgments}
P.M. Shepley is greatful to cryogen technicians Luke Bone, Brian Gibbs, Dave Chapman and John Turton for a steady supply of liquid Helium, and to Dr Mannan Ali and Jamie Massey for cryostat advice. The authors acknowledge funding from a CASE studentship from EPSRC and Thales UK (L.A. Stoica), the School of Chemical and Process Engineering (P.M. Shepley) and EPSRC award EP/M002462/1 (A.J. Bell).
\end{acknowledgments}

%\newpage{}

% If in two-column mode, this environment will change to single-column format so that long equations can be displayed. 
%\begin{widetext}
%$$\mbox{put long equation here}$$
%\end{widetext}

% Figures should be put into the text as floats. 
% Use the graphics or graphicx packages (distributed with LaTeX2e).
% Put the label that you will use with \ref{} command in the braces of the \label{} command.
%
% \begin{figure}
% \end{figure}

% \begin{table}
% \end{tabular}
% \end{table}

%\begin{acknowledgments}
% Put your acknowledgments here.
%\end{acknowledgments}

\bibliography{refs_PINPMNPT_PMShepley2018}

\begin{thebibliography}{30}%
\makeatletter
\providecommand \@ifxundefined [1]{%
 \@ifx{#1\undefined}
}%
\providecommand \@ifnum [1]{%
 \ifnum #1\expandafter \@firstoftwo
 \else \expandafter \@secondoftwo
 \fi
}%
\providecommand \@ifx [1]{%
 \ifx #1\expandafter \@firstoftwo
 \else \expandafter \@secondoftwo
 \fi
}%
\providecommand \natexlab [1]{#1}%
\providecommand \enquote  [1]{``#1''}%
\providecommand \bibnamefont  [1]{#1}%
\providecommand \bibfnamefont [1]{#1}%
\providecommand \citenamefont [1]{#1}%
\providecommand \href@noop [0]{\@secondoftwo}%
\providecommand \href [0]{\begingroup \@sanitize@url \@href}%
\providecommand \@href[1]{\@@startlink{#1}\@@href}%
\providecommand \@@href[1]{\endgroup#1\@@endlink}%
\providecommand \@sanitize@url [0]{\catcode `\\12\catcode `\$12\catcode
  `\&12\catcode `\#12\catcode `\^12\catcode `\_12\catcode `\%12\relax}%
\providecommand \@@startlink[1]{}%
\providecommand \@@endlink[0]{}%
\providecommand \url  [0]{\begingroup\@sanitize@url \@url }%
\providecommand \@url [1]{\endgroup\@href {#1}{\urlprefix }}%
\providecommand \urlprefix  [0]{URL }%
\providecommand \Eprint [0]{\href }%
\providecommand \doibase [0]{http://dx.doi.org/}%
\providecommand \selectlanguage [0]{\@gobble}%
\providecommand \bibinfo  [0]{\@secondoftwo}%
\providecommand \bibfield  [0]{\@secondoftwo}%
\providecommand \translation [1]{[#1]}%
\providecommand \BibitemOpen [0]{}%
\providecommand \bibitemStop [0]{}%
\providecommand \bibitemNoStop [0]{.\EOS\space}%
\providecommand \EOS [0]{\spacefactor3000\relax}%
\providecommand \BibitemShut  [1]{\csname bibitem#1\endcsname}%
\let\auto@bib@innerbib\@empty
%</preamble>
\bibitem [{\citenamefont {Park}\ and\ \citenamefont {Shrout}(1997)}]{Park1997}%
  \BibitemOpen
  \bibfield  {author} {\bibinfo {author} {\bibfnamefont {S.-E.}\ \bibnamefont
  {Park}}\ and\ \bibinfo {author} {\bibfnamefont {T.~R.}\ \bibnamefont
  {Shrout}},\ }\bibfield  {title} {\enquote {\bibinfo {title} {{Ultrahigh
  strain and piezoelectric behavior in relaxor based ferroelectric single
  crystals}},}\ }\href {\doibase 10.1063/1.365983} {\bibfield  {journal}
  {\bibinfo  {journal} {Journal of Applied Physics}\ }\textbf {\bibinfo
  {volume} {82}},\ \bibinfo {pages} {1804--1811} (\bibinfo {year}
  {1997})}\BibitemShut {NoStop}%
\bibitem [{\citenamefont {Zhang}\ and\ \citenamefont {Li}(2012)}]{Zhang2012}%
  \BibitemOpen
  \bibfield  {author} {\bibinfo {author} {\bibfnamefont {S.}~\bibnamefont
  {Zhang}}\ and\ \bibinfo {author} {\bibfnamefont {F.}~\bibnamefont {Li}},\
  }\href {\doibase 10.1063/1.3679521} {\enquote {\bibinfo {title} {{High
  performance ferroelectric relaxor-PbTiO 3 single crystals: Status and
  perspective}},}\ } (\bibinfo {year} {2012})\BibitemShut {NoStop}%
\bibitem [{\citenamefont {Zhang}\ \emph {et~al.}(2015)\citenamefont {Zhang},
  \citenamefont {Li}, \citenamefont {Jiang}, \citenamefont {Kim}, \citenamefont
  {Luo},\ and\ \citenamefont {Geng}}]{Zhang2015}%
  \BibitemOpen
  \bibfield  {author} {\bibinfo {author} {\bibfnamefont {S.}~\bibnamefont
  {Zhang}}, \bibinfo {author} {\bibfnamefont {F.}~\bibnamefont {Li}}, \bibinfo
  {author} {\bibfnamefont {X.}~\bibnamefont {Jiang}}, \bibinfo {author}
  {\bibfnamefont {J.}~\bibnamefont {Kim}}, \bibinfo {author} {\bibfnamefont
  {J.}~\bibnamefont {Luo}}, \ and\ \bibinfo {author} {\bibfnamefont
  {X.}~\bibnamefont {Geng}},\ }\bibfield  {title} {\enquote {\bibinfo {title}
  {{Advantages and challenges of relaxor-PbTiO3 ferroelectric crystals for
  electroacoustic transducers - A review}},}\ }\href {\doibase
  10.1016/j.pmatsci.2014.10.002} {\bibfield  {journal} {\bibinfo  {journal}
  {Progress in Materials Science}\ }\textbf {\bibinfo {volume} {68}},\ \bibinfo
  {pages} {1--66} (\bibinfo {year} {2015})}\BibitemShut {NoStop}%
\bibitem [{\citenamefont {Martin}\ \emph {et~al.}(2012)\citenamefont {Martin},
  \citenamefont {ter Brake}, \citenamefont {Lebrun}, \citenamefont {Zhang},\
  and\ \citenamefont {Shrout}}]{Martin2012}%
  \BibitemOpen
  \bibfield  {author} {\bibinfo {author} {\bibfnamefont {F.}~\bibnamefont
  {Martin}}, \bibinfo {author} {\bibfnamefont {H.~J.~M.}\ \bibnamefont {ter
  Brake}}, \bibinfo {author} {\bibfnamefont {L.}~\bibnamefont {Lebrun}},
  \bibinfo {author} {\bibfnamefont {S.}~\bibnamefont {Zhang}}, \ and\ \bibinfo
  {author} {\bibfnamefont {T.}~\bibnamefont {Shrout}},\ }\bibfield  {title}
  {\enquote {\bibinfo {title} {{Dielectric and piezoelectric activities in
  (1− x )Pb(Mg 1/3 Nb 2/3 )O 3 − x PbTiO 3 single crystals from 5 K to 300
  K}},}\ }\href {\doibase 10.1063/1.4716031} {\bibfield  {journal} {\bibinfo
  {journal} {Journal of Applied Physics}\ }\textbf {\bibinfo {volume} {111}},\
  \bibinfo {pages} {104108} (\bibinfo {year} {2012})}\BibitemShut {NoStop}%
\bibitem [{\citenamefont {Bukhari}\ \emph {et~al.}(2014)\citenamefont
  {Bukhari}, \citenamefont {Islam}, \citenamefont {Haziot},\ and\ \citenamefont
  {Beamish}}]{Bukhari2014}%
  \BibitemOpen
  \bibfield  {author} {\bibinfo {author} {\bibfnamefont {S.}~\bibnamefont
  {Bukhari}}, \bibinfo {author} {\bibfnamefont {M.}~\bibnamefont {Islam}},
  \bibinfo {author} {\bibfnamefont {A.}~\bibnamefont {Haziot}}, \ and\ \bibinfo
  {author} {\bibfnamefont {J.}~\bibnamefont {Beamish}},\ }\bibfield  {title}
  {\enquote {\bibinfo {title} {{Shear piezoelectric coefficients of PZT, LiNbO
  3 and PMN-PT at cryogenic temperatures}},}\ }\href {\doibase
  10.1088/1742-6596/568/3/032004} {\bibfield  {journal} {\bibinfo  {journal}
  {Journal of Physics: Conference Series}\ }\textbf {\bibinfo {volume} {568}},\
  \bibinfo {pages} {032004} (\bibinfo {year} {2014})}\BibitemShut {NoStop}%
\bibitem [{\citenamefont {Li}\ \emph {et~al.}(2016)\citenamefont {Li},
  \citenamefont {Zhang}, \citenamefont {Yang}, \citenamefont {Xu},
  \citenamefont {Zhang}, \citenamefont {Liu}, \citenamefont {Wang},
  \citenamefont {Wang}, \citenamefont {Cheng}, \citenamefont {Ye},
  \citenamefont {Luo}, \citenamefont {Shrout},\ and\ \citenamefont
  {Chen}}]{Li2016}%
  \BibitemOpen
  \bibfield  {author} {\bibinfo {author} {\bibfnamefont {F.}~\bibnamefont
  {Li}}, \bibinfo {author} {\bibfnamefont {S.}~\bibnamefont {Zhang}}, \bibinfo
  {author} {\bibfnamefont {T.}~\bibnamefont {Yang}}, \bibinfo {author}
  {\bibfnamefont {Z.}~\bibnamefont {Xu}}, \bibinfo {author} {\bibfnamefont
  {N.}~\bibnamefont {Zhang}}, \bibinfo {author} {\bibfnamefont
  {G.}~\bibnamefont {Liu}}, \bibinfo {author} {\bibfnamefont {J.}~\bibnamefont
  {Wang}}, \bibinfo {author} {\bibfnamefont {J.}~\bibnamefont {Wang}}, \bibinfo
  {author} {\bibfnamefont {Z.}~\bibnamefont {Cheng}}, \bibinfo {author}
  {\bibfnamefont {Z.-G.}\ \bibnamefont {Ye}}, \bibinfo {author} {\bibfnamefont
  {J.}~\bibnamefont {Luo}}, \bibinfo {author} {\bibfnamefont {T.~R.}\
  \bibnamefont {Shrout}}, \ and\ \bibinfo {author} {\bibfnamefont {L.-Q.}\
  \bibnamefont {Chen}},\ }\bibfield  {title} {\enquote {\bibinfo {title} {{The
  origin of ultrahigh piezoelectricity in relaxor-ferroelectric solid solution
  crystals}},}\ }\href {\doibase 10.1038/ncomms13807} {\bibfield  {journal}
  {\bibinfo  {journal} {Nature Communications}\ }\textbf {\bibinfo {volume}
  {7}},\ \bibinfo {pages} {13807} (\bibinfo {year} {2016})}\BibitemShut
  {NoStop}%
\bibitem [{\citenamefont {Wang}\ \emph {et~al.}(2009)\citenamefont {Wang},
  \citenamefont {Or}, \citenamefont {Zhao},\ and\ \citenamefont
  {Luo}}]{Wang2009}%
  \BibitemOpen
  \bibfield  {author} {\bibinfo {author} {\bibfnamefont {F.}~\bibnamefont
  {Wang}}, \bibinfo {author} {\bibfnamefont {S.~W.}\ \bibnamefont {Or}},
  \bibinfo {author} {\bibfnamefont {X.}~\bibnamefont {Zhao}}, \ and\ \bibinfo
  {author} {\bibfnamefont {H.}~\bibnamefont {Luo}},\ }\bibfield  {title}
  {\enquote {\bibinfo {title} {{Cryogenic dielectric and piezoelectric
  activities in rhombohedral (1 − x )Pb(Mg 1/3 Nb 2/3 )O 3 – x PbTiO 3
  single crystals with different crystallographic orientations}},}\ }\href
  {\doibase 10.1088/0022-3727/42/18/182001} {\bibfield  {journal} {\bibinfo
  {journal} {Journal of Physics D: Applied Physics}\ }\textbf {\bibinfo
  {volume} {42}},\ \bibinfo {pages} {182001} (\bibinfo {year}
  {2009})}\BibitemShut {NoStop}%
\bibitem [{\citenamefont {Wang}\ \emph {et~al.}(2014)\citenamefont {Wang},
  \citenamefont {Wang}, \citenamefont {Luo}, \citenamefont {Li},\ and\
  \citenamefont {Viehland}}]{Wang2014}%
  \BibitemOpen
  \bibfield  {author} {\bibinfo {author} {\bibfnamefont {Z.}~\bibnamefont
  {Wang}}, \bibinfo {author} {\bibfnamefont {Y.}~\bibnamefont {Wang}}, \bibinfo
  {author} {\bibfnamefont {H.}~\bibnamefont {Luo}}, \bibinfo {author}
  {\bibfnamefont {J.}~\bibnamefont {Li}}, \ and\ \bibinfo {author}
  {\bibfnamefont {D.}~\bibnamefont {Viehland}},\ }\bibfield  {title} {\enquote
  {\bibinfo {title} {{Crafting the strain state in epitaxial thin films: A case
  study of CoFe2O4 films on Pb(Mg,Nb)O3-PbTiO3}},}\ }\href {\doibase
  10.1103/PhysRevB.90.134103} {\bibfield  {journal} {\bibinfo  {journal}
  {Physical Review B}\ }\textbf {\bibinfo {volume} {90}},\ \bibinfo {pages}
  {134103} (\bibinfo {year} {2014})}\BibitemShut {NoStop}%
\bibitem [{\citenamefont {Takenaka}\ \emph {et~al.}(2017)\citenamefont
  {Takenaka}, \citenamefont {Grinberg}, \citenamefont {Liu},\ and\
  \citenamefont {Rappe}}]{Takenaka2017}%
  \BibitemOpen
  \bibfield  {author} {\bibinfo {author} {\bibfnamefont {H.}~\bibnamefont
  {Takenaka}}, \bibinfo {author} {\bibfnamefont {I.}~\bibnamefont {Grinberg}},
  \bibinfo {author} {\bibfnamefont {S.}~\bibnamefont {Liu}}, \ and\ \bibinfo
  {author} {\bibfnamefont {A.~M.}\ \bibnamefont {Rappe}},\ }\bibfield  {title}
  {\enquote {\bibinfo {title} {{Slush-like polar structures in single-crystal
  relaxors}},}\ }\href {\doibase 10.1038/nature22068} {\bibfield  {journal}
  {\bibinfo  {journal} {Nature}\ }\textbf {\bibinfo {volume} {546}},\ \bibinfo
  {pages} {391--395} (\bibinfo {year} {2017})}\BibitemShut {NoStop}%
\bibitem [{\citenamefont {Jin}, \citenamefont {Porokhonskyy},\ and\
  \citenamefont {Damjanovic}(2010)}]{Porokhonskyy2010}%
  \BibitemOpen
  \bibfield  {author} {\bibinfo {author} {\bibfnamefont {L.}~\bibnamefont
  {Jin}}, \bibinfo {author} {\bibfnamefont {V.}~\bibnamefont {Porokhonskyy}}, \
  and\ \bibinfo {author} {\bibfnamefont {D.}~\bibnamefont {Damjanovic}},\
  }\bibfield  {title} {\enquote {\bibinfo {title} {{Domain wall contributions
  in Pb(Zr,Ti)O3 ceramics at morphotropic phase boundary: A study of dielectric
  dispersion}},}\ }\href {\doibase 10.1063/1.3455328} {\bibfield  {journal}
  {\bibinfo  {journal} {Applied Physics Letters}\ }\textbf {\bibinfo {volume}
  {96}},\ \bibinfo {pages} {242902} (\bibinfo {year} {2010})}\BibitemShut
  {NoStop}%
\bibitem [{\citenamefont {Li}\ \emph {et~al.}(2017)\citenamefont {Li},
  \citenamefont {Zhang}, \citenamefont {Xu},\ and\ \citenamefont
  {Chen}}]{Li2017}%
  \BibitemOpen
  \bibfield  {author} {\bibinfo {author} {\bibfnamefont {F.}~\bibnamefont
  {Li}}, \bibinfo {author} {\bibfnamefont {S.}~\bibnamefont {Zhang}}, \bibinfo
  {author} {\bibfnamefont {Z.}~\bibnamefont {Xu}}, \ and\ \bibinfo {author}
  {\bibfnamefont {L.-Q.}\ \bibnamefont {Chen}},\ }\bibfield  {title} {\enquote
  {\bibinfo {title} {{The Contributions of Polar Nanoregions to the Dielectric
  and Piezoelectric Responses in Domain-Engineered Relaxor-PbTiO 3
  Crystals}},}\ }\href {\doibase 10.1002/adfm.201700310} {\bibfield  {journal}
  {\bibinfo  {journal} {Advanced Functional Materials}\ }\textbf {\bibinfo
  {volume} {27}},\ \bibinfo {pages} {1700310} (\bibinfo {year}
  {2017})}\BibitemShut {NoStop}%
\bibitem [{\citenamefont {Lente}\ \emph {et~al.}(2004)\citenamefont {Lente},
  \citenamefont {Zanin}, \citenamefont {Andreeta}, \citenamefont {Santos},
  \citenamefont {Garcia},\ and\ \citenamefont {Eiras}}]{Lente2004}%
  \BibitemOpen
  \bibfield  {author} {\bibinfo {author} {\bibfnamefont {M.~H.}\ \bibnamefont
  {Lente}}, \bibinfo {author} {\bibfnamefont {A.~L.}\ \bibnamefont {Zanin}},
  \bibinfo {author} {\bibfnamefont {E.~R.~M.}\ \bibnamefont {Andreeta}},
  \bibinfo {author} {\bibfnamefont {I.~A.}\ \bibnamefont {Santos}}, \bibinfo
  {author} {\bibfnamefont {D.}~\bibnamefont {Garcia}}, \ and\ \bibinfo {author}
  {\bibfnamefont {J.~A.}\ \bibnamefont {Eiras}},\ }\bibfield  {title} {\enquote
  {\bibinfo {title} {{Investigation of dielectric anomalies at cryogenic
  temperatures in (1-x)[Pb(Mg1/3Nb2/3)O3]-xPbTiO3 system}},}\ }\href {\doibase
  10.1063/1.1771807} {\bibfield  {journal} {\bibinfo  {journal} {Applied
  Physics Letters}\ }\textbf {\bibinfo {volume} {85}},\ \bibinfo {pages}
  {982--984} (\bibinfo {year} {2004})},\ \Eprint {http://arxiv.org/abs/0407042}
  {arXiv:0407042 [cond-mat]} \BibitemShut {NoStop}%
\bibitem [{\citenamefont {Hentati}\ \emph {et~al.}(2015)\citenamefont
  {Hentati}, \citenamefont {Dammak}, \citenamefont {Khemakhem}, \citenamefont
  {Guiblin},\ and\ \citenamefont {Thi}}]{Hentati2015}%
  \BibitemOpen
  \bibfield  {author} {\bibinfo {author} {\bibfnamefont {M.~A.}\ \bibnamefont
  {Hentati}}, \bibinfo {author} {\bibfnamefont {H.}~\bibnamefont {Dammak}},
  \bibinfo {author} {\bibfnamefont {H.}~\bibnamefont {Khemakhem}}, \bibinfo
  {author} {\bibfnamefont {N.}~\bibnamefont {Guiblin}}, \ and\ \bibinfo
  {author} {\bibfnamefont {M.~P.}\ \bibnamefont {Thi}},\ }\bibfield  {title}
  {\enquote {\bibinfo {title} {{Dielectric evidence of persistence of polar
  nanoregions within the ferroelectric phases of (1-x)Pb(Zn1/3Nb2/3)O3-xPbTiO3
  relaxor ferroelectric system}},}\ }\href {\doibase 10.1063/1.4926877}
  {\bibfield  {journal} {\bibinfo  {journal} {Journal of Applied Physics}\
  }\textbf {\bibinfo {volume} {118}},\ \bibinfo {pages} {034104} (\bibinfo
  {year} {2015})}\BibitemShut {NoStop}%
\bibitem [{\citenamefont {Singh}\ \emph {et~al.}(2007)\citenamefont {Singh},
  \citenamefont {Singh}, \citenamefont {Pandey},\ and\ \citenamefont
  {Yusuf}}]{Singh2007}%
  \BibitemOpen
  \bibfield  {author} {\bibinfo {author} {\bibfnamefont {S.~P.}\ \bibnamefont
  {Singh}}, \bibinfo {author} {\bibfnamefont {A.~K.}\ \bibnamefont {Singh}},
  \bibinfo {author} {\bibfnamefont {D.}~\bibnamefont {Pandey}}, \ and\ \bibinfo
  {author} {\bibfnamefont {S.~M.}\ \bibnamefont {Yusuf}},\ }\bibfield  {title}
  {\enquote {\bibinfo {title} {{Dielectric relaxation and phase transitions at
  cryogenic temperatures in 0.65 [Pb (Ni13 Nb23) O3] -0.35PbTi O3 ceramics}},}\
  }\href {\doibase 10.1103/PhysRevB.76.054102} {\bibfield  {journal} {\bibinfo
  {journal} {Physical Review B - Condensed Matter and Materials Physics}\
  }\textbf {\bibinfo {volume} {76}},\ \bibinfo {pages} {1--8} (\bibinfo {year}
  {2007})}\BibitemShut {NoStop}%
\bibitem [{\citenamefont {Li}\ \emph {et~al.}(2008)\citenamefont {Li},
  \citenamefont {Xu}, \citenamefont {Xi}, \citenamefont {Cao},\ and\
  \citenamefont {Yao}}]{Li2008}%
  \BibitemOpen
  \bibfield  {author} {\bibinfo {author} {\bibfnamefont {Z.}~\bibnamefont
  {Li}}, \bibinfo {author} {\bibfnamefont {Z.}~\bibnamefont {Xu}}, \bibinfo
  {author} {\bibfnamefont {Z.}~\bibnamefont {Xi}}, \bibinfo {author}
  {\bibfnamefont {L.}~\bibnamefont {Cao}}, \ and\ \bibinfo {author}
  {\bibfnamefont {X.}~\bibnamefont {Yao}},\ }\bibfield  {title} {\enquote
  {\bibinfo {title} {{Dielectric loss anomalies of 0.68PMN-0.32PT single
  crystal and ceramics at cryogenic temperature}},}\ }\href {\doibase
  10.1007/s10832-007-9150-2} {\bibfield  {journal} {\bibinfo  {journal}
  {Journal of Electroceramics}\ }\textbf {\bibinfo {volume} {21}},\ \bibinfo
  {pages} {279--282} (\bibinfo {year} {2008})}\BibitemShut {NoStop}%
\bibitem [{\citenamefont {Zhang}\ \emph {et~al.}(1983)\citenamefont {Zhang},
  \citenamefont {Chen}, \citenamefont {Cross},\ and\ \citenamefont
  {Schulze}}]{Zhang1983}%
  \BibitemOpen
  \bibfield  {author} {\bibinfo {author} {\bibfnamefont {X.~L.}\ \bibnamefont
  {Zhang}}, \bibinfo {author} {\bibfnamefont {Z.~X.}\ \bibnamefont {Chen}},
  \bibinfo {author} {\bibfnamefont {L.~E.}\ \bibnamefont {Cross}}, \ and\
  \bibinfo {author} {\bibfnamefont {W.~A.}\ \bibnamefont {Schulze}},\
  }\bibfield  {title} {\enquote {\bibinfo {title} {{Dielectric and
  piezoelectric properties of modified lead titanate zirconate ceramics from
  4.2 to 300 K}},}\ }\href {\doibase 10.1007/BF00551962} {\bibfield  {journal}
  {\bibinfo  {journal} {Journal of Materials Science}\ }\textbf {\bibinfo
  {volume} {18}},\ \bibinfo {pages} {968--972} (\bibinfo {year}
  {1983})}\BibitemShut {NoStop}%
\bibitem [{\citenamefont {Zhang}\ \emph {et~al.}(1994)\citenamefont {Zhang},
  \citenamefont {Wang}, \citenamefont {Kim},\ and\ \citenamefont
  {Cross}}]{Zhang1994}%
  \BibitemOpen
  \bibfield  {author} {\bibinfo {author} {\bibfnamefont {Q.~M.}\ \bibnamefont
  {Zhang}}, \bibinfo {author} {\bibfnamefont {H.}~\bibnamefont {Wang}},
  \bibinfo {author} {\bibfnamefont {N.}~\bibnamefont {Kim}}, \ and\ \bibinfo
  {author} {\bibfnamefont {L.~E.}\ \bibnamefont {Cross}},\ }\bibfield  {title}
  {\enquote {\bibinfo {title} {{Direct evaluation of domain‐wall and
  intrinsic contributions to the dielectric and piezoelectric response and
  their temperature dependence on lead zirconate‐titanate ceramics}},}\
  }\href {\doibase 10.1063/1.355874} {\bibfield  {journal} {\bibinfo  {journal}
  {Journal of Applied Physics}\ }\textbf {\bibinfo {volume} {75}},\ \bibinfo
  {pages} {454--459} (\bibinfo {year} {1994})}\BibitemShut {NoStop}%
\bibitem [{\citenamefont {Thiercelin}, \citenamefont {Dammak},\ and\
  \citenamefont {{Pham Thi}}(2010)}]{Thiercelin2010}%
  \BibitemOpen
  \bibfield  {author} {\bibinfo {author} {\bibfnamefont {M.}~\bibnamefont
  {Thiercelin}}, \bibinfo {author} {\bibfnamefont {H.}~\bibnamefont {Dammak}},
  \ and\ \bibinfo {author} {\bibfnamefont {M.}~\bibnamefont {{Pham Thi}}},\
  }\bibfield  {title} {\enquote {\bibinfo {title} {{Electromechanical
  properties of PMN-PT and PZT ceramics at cryogenic temperatures}},}\ }\href
  {\doibase 10.1109/ISAF.2010.5712258} {\bibfield  {journal} {\bibinfo
  {journal} {Proceedings of the 2010 IEEE International Symposium on the
  Applications of Ferroelectrics, ISAF 2010, Co-located with the 10th European
  Conference on the Applications of Polar Dielectrics, ECAPD 2010}\ ,\ \bibinfo
  {pages} {2--5}} (\bibinfo {year} {2010})}\BibitemShut {NoStop}%
\bibitem [{\citenamefont {Li}\ \emph {et~al.}(2014)\citenamefont {Li},
  \citenamefont {Wang}, \citenamefont {Jin}, \citenamefont {Xu},\ and\
  \citenamefont {Zhang}}]{Li2014}%
  \BibitemOpen
  \bibfield  {author} {\bibinfo {author} {\bibfnamefont {F.}~\bibnamefont
  {Li}}, \bibinfo {author} {\bibfnamefont {L.}~\bibnamefont {Wang}}, \bibinfo
  {author} {\bibfnamefont {L.}~\bibnamefont {Jin}}, \bibinfo {author}
  {\bibfnamefont {Z.}~\bibnamefont {Xu}}, \ and\ \bibinfo {author}
  {\bibfnamefont {S.}~\bibnamefont {Zhang}},\ }\bibfield  {title} {\enquote
  {\bibinfo {title} {{Achieving single domain relaxor-PT crystals by high
  temperature poling}},}\ }\href {\doibase 10.1039/c3ce42330a} {\bibfield
  {journal} {\bibinfo  {journal} {CrystEngComm}\ }\textbf {\bibinfo {volume}
  {16}},\ \bibinfo {pages} {2892--2897} (\bibinfo {year} {2014})}\BibitemShut
  {NoStop}%
\bibitem [{\citenamefont {Arlt}, \citenamefont {Dederichs},\ and\ \citenamefont
  {Herbiet}(1987)}]{Arlt1987}%
  \BibitemOpen
  \bibfield  {author} {\bibinfo {author} {\bibfnamefont {G.}~\bibnamefont
  {Arlt}}, \bibinfo {author} {\bibfnamefont {H.}~\bibnamefont {Dederichs}}, \
  and\ \bibinfo {author} {\bibfnamefont {R.}~\bibnamefont {Herbiet}},\
  }\bibfield  {title} {\enquote {\bibinfo {title} {90°-domain wall relaxation
  in tetragonally distorted ferroelectric ceramics},}\ }\href {\doibase
  10.1080/00150198708014493} {\bibfield  {journal} {\bibinfo  {journal}
  {Ferroelectrics}\ }\textbf {\bibinfo {volume} {74}},\ \bibinfo {pages}
  {37--53} (\bibinfo {year} {1987})}\BibitemShut {NoStop}%
\bibitem [{\citenamefont {Akishige}, \citenamefont {Nakanishi},\ and\
  \citenamefont {Mōri}(1998)}]{Akishige1998}%
  \BibitemOpen
  \bibfield  {author} {\bibinfo {author} {\bibfnamefont {Y.}~\bibnamefont
  {Akishige}}, \bibinfo {author} {\bibfnamefont {T.}~\bibnamefont {Nakanishi}},
  \ and\ \bibinfo {author} {\bibfnamefont {N.}~\bibnamefont {Mōri}},\
  }\bibfield  {title} {\enquote {\bibinfo {title} {{Dielectric dispersion in
  BaTiO 3 single crystal at low temperatures}},}\ }\href {\doibase
  10.1080/00150199808015041} {\bibfield  {journal} {\bibinfo  {journal}
  {Ferroelectrics}\ }\textbf {\bibinfo {volume} {217}},\ \bibinfo {pages}
  {217--222} (\bibinfo {year} {1998})}\BibitemShut {NoStop}%
\bibitem [{\citenamefont {Akishige}, \citenamefont {Fukano},\ and\
  \citenamefont {Shigematsu}(2004)}]{Akishige2004}%
  \BibitemOpen
  \bibfield  {author} {\bibinfo {author} {\bibfnamefont {Y.}~\bibnamefont
  {Akishige}}, \bibinfo {author} {\bibfnamefont {K.}~\bibnamefont {Fukano}}, \
  and\ \bibinfo {author} {\bibfnamefont {H.}~\bibnamefont {Shigematsu}},\
  }\bibfield  {title} {\enquote {\bibinfo {title} {{Crystal Growth and
  Dielectric Properties of New Ferroelectric Barium Titanate: BaTi2O5}},}\
  }\href {\doibase 10.1007/s10832-004-5158-z} {\bibfield  {journal} {\bibinfo
  {journal} {Journal of Electroceramics}\ }\textbf {\bibinfo {volume} {13}},\
  \bibinfo {pages} {561--565} (\bibinfo {year} {2004})}\BibitemShut {NoStop}%
\bibitem [{\citenamefont {Wang}\ \emph {et~al.}(2007)\citenamefont {Wang},
  \citenamefont {Tagantsev}, \citenamefont {Damjanovic},\ and\ \citenamefont
  {Setter}}]{Wang2007}%
  \BibitemOpen
  \bibfield  {author} {\bibinfo {author} {\bibfnamefont {Y.~L.}\ \bibnamefont
  {Wang}}, \bibinfo {author} {\bibfnamefont {A.~K.}\ \bibnamefont {Tagantsev}},
  \bibinfo {author} {\bibfnamefont {D.}~\bibnamefont {Damjanovic}}, \ and\
  \bibinfo {author} {\bibfnamefont {N.}~\bibnamefont {Setter}},\ }\bibfield
  {title} {\enquote {\bibinfo {title} {{Giant domain wall contribution to the
  dielectric susceptibility in BaTiO3 single crystals}},}\ }\href {\doibase
  10.1063/1.2751135} {\bibfield  {journal} {\bibinfo  {journal} {Applied
  Physics Letters}\ }\textbf {\bibinfo {volume} {91}},\ \bibinfo {pages}
  {062905} (\bibinfo {year} {2007})}\BibitemShut {NoStop}%
\bibitem [{\citenamefont {Wang}\ \emph {et~al.}(2011)\citenamefont {Wang},
  \citenamefont {He}, \citenamefont {Damjanovic}, \citenamefont {Tagantsev},
  \citenamefont {Deng},\ and\ \citenamefont {Setter}}]{Wang2011}%
  \BibitemOpen
  \bibfield  {author} {\bibinfo {author} {\bibfnamefont {Y.~L.}\ \bibnamefont
  {Wang}}, \bibinfo {author} {\bibfnamefont {Z.~B.}\ \bibnamefont {He}},
  \bibinfo {author} {\bibfnamefont {D.}~\bibnamefont {Damjanovic}}, \bibinfo
  {author} {\bibfnamefont {A.~K.}\ \bibnamefont {Tagantsev}}, \bibinfo {author}
  {\bibfnamefont {G.~C.}\ \bibnamefont {Deng}}, \ and\ \bibinfo {author}
  {\bibfnamefont {N.}~\bibnamefont {Setter}},\ }\bibfield  {title} {\enquote
  {\bibinfo {title} {{Unusual dielectric behavior and domain structure in
  rhombohedral phase of BaTiO 3 single crystals}},}\ }\href {\doibase
  10.1063/1.3605494} {\bibfield  {journal} {\bibinfo  {journal} {Journal of
  Applied Physics}\ }\textbf {\bibinfo {volume} {110}},\ \bibinfo {pages}
  {014101} (\bibinfo {year} {2011})}\BibitemShut {NoStop}%
\bibitem [{\citenamefont {Stoica}(2016)}]{Stoica2016}%
  \BibitemOpen
  \bibfield  {author} {\bibinfo {author} {\bibfnamefont {L.~A.}\ \bibnamefont
  {Stoica}},\ }\emph {\bibinfo {title} {{Relaxor-PbTiO3 single crystals and
  polycrystals: processing, growth and characterisation}}},\ \href
  {http://etheses.whiterose.ac.uk/id/eprint/16259} {\bibinfo {type} {Thesis
  (phd)}},\ \bibinfo  {school} {University of Leeds} (\bibinfo {year}
  {2016})\BibitemShut {NoStop}%
\bibitem [{\citenamefont {Kiat}\ \emph {et~al.}(2002)\citenamefont {Kiat},
  \citenamefont {Uesu}, \citenamefont {Dkhil}, \citenamefont {Matsuda},
  \citenamefont {Malibert},\ and\ \citenamefont {Calvarin}}]{Kiat2002}%
  \BibitemOpen
  \bibfield  {author} {\bibinfo {author} {\bibfnamefont {J.-M.}\ \bibnamefont
  {Kiat}}, \bibinfo {author} {\bibfnamefont {Y.}~\bibnamefont {Uesu}}, \bibinfo
  {author} {\bibfnamefont {B.}~\bibnamefont {Dkhil}}, \bibinfo {author}
  {\bibfnamefont {M.}~\bibnamefont {Matsuda}}, \bibinfo {author} {\bibfnamefont
  {C.}~\bibnamefont {Malibert}}, \ and\ \bibinfo {author} {\bibfnamefont
  {G.}~\bibnamefont {Calvarin}},\ }\bibfield  {title} {\enquote {\bibinfo
  {title} {{Monoclinic structure of unpoled morphotropic high piezoelectric
  PMN-PT and PZN-PT compounds}},}\ }\href {\doibase 10.1103/PhysRevB.65.064106}
  {\bibfield  {journal} {\bibinfo  {journal} {Physical Review B}\ }\textbf
  {\bibinfo {volume} {65}},\ \bibinfo {pages} {064106} (\bibinfo {year}
  {2002})},\ \Eprint {http://arxiv.org/abs/0109217} {arXiv:0109217 [cond-mat]}
  \BibitemShut {NoStop}%
\bibitem [{\citenamefont {Cao}\ \emph {et~al.}(2006)\citenamefont {Cao},
  \citenamefont {Li}, \citenamefont {Viehland},\ and\ \citenamefont
  {Xu}}]{Cao2006a}%
  \BibitemOpen
  \bibfield  {author} {\bibinfo {author} {\bibfnamefont {H.}~\bibnamefont
  {Cao}}, \bibinfo {author} {\bibfnamefont {J.}~\bibnamefont {Li}}, \bibinfo
  {author} {\bibfnamefont {D.}~\bibnamefont {Viehland}}, \ and\ \bibinfo
  {author} {\bibfnamefont {G.}~\bibnamefont {Xu}},\ }\bibfield  {title}
  {\enquote {\bibinfo {title} {{Fragile phase stability in
  „1−x{\ldots}Pb„Mg1/3Nb2/3O3{\ldots}-xPbTiO3 crystals: A comparison of
  [001] and [110] field-cooled phase diagrams}},}\ }\href {\doibase
  10.1103/PhysRevB.73.184110} {\bibfield  {journal} {\bibinfo  {journal}
  {Physical Review B}\ }\textbf {\bibinfo {volume} {73}},\ \bibinfo {pages}
  {184110} (\bibinfo {year} {2006})}\BibitemShut {NoStop}%
\bibitem [{\citenamefont {Shvartsman}\ and\ \citenamefont
  {Kholkin}(2004)}]{Shvartsman2004}%
  \BibitemOpen
  \bibfield  {author} {\bibinfo {author} {\bibfnamefont {V.~V.}\ \bibnamefont
  {Shvartsman}}\ and\ \bibinfo {author} {\bibfnamefont {A.~L.}\ \bibnamefont
  {Kholkin}},\ }\bibfield  {title} {\enquote {\bibinfo {title} {{Domain
  structure of 0.8Pb(Mg1/3Nb2/3)O3-0.2PbTiO3 studied by piezoresponse force
  microscopy}},}\ }\href {\doibase 10.1103/PhysRevB.69.014102} {\bibfield
  {journal} {\bibinfo  {journal} {Physical Review B}\ }\textbf {\bibinfo
  {volume} {69}},\ \bibinfo {pages} {014102} (\bibinfo {year}
  {2004})}\BibitemShut {NoStop}%
\bibitem [{\citenamefont {Sch{\"{o}}nau}\ \emph {et~al.}(2007)\citenamefont
  {Sch{\"{o}}nau}, \citenamefont {Schmitt}, \citenamefont {Knapp},
  \citenamefont {Fuess}, \citenamefont {Eichel}, \citenamefont {Kungl},\ and\
  \citenamefont {Hoffmann}}]{Schonau2007}%
  \BibitemOpen
  \bibfield  {author} {\bibinfo {author} {\bibfnamefont {K.~A.}\ \bibnamefont
  {Sch{\"{o}}nau}}, \bibinfo {author} {\bibfnamefont {L.~A.}\ \bibnamefont
  {Schmitt}}, \bibinfo {author} {\bibfnamefont {M.}~\bibnamefont {Knapp}},
  \bibinfo {author} {\bibfnamefont {H.}~\bibnamefont {Fuess}}, \bibinfo
  {author} {\bibfnamefont {R.-A.}\ \bibnamefont {Eichel}}, \bibinfo {author}
  {\bibfnamefont {H.}~\bibnamefont {Kungl}}, \ and\ \bibinfo {author}
  {\bibfnamefont {M.~J.}\ \bibnamefont {Hoffmann}},\ }\bibfield  {title}
  {\enquote {\bibinfo {title} {{Nanodomain structure of Pb[Zr1−xTix]O3 at its
  morphotropic phase boundary: Investigations from local to average
  structure}},}\ }\href {\doibase 10.1103/PhysRevB.75.184117} {\bibfield
  {journal} {\bibinfo  {journal} {Physical Review B}\ }\textbf {\bibinfo
  {volume} {75}},\ \bibinfo {pages} {184117} (\bibinfo {year}
  {2007})}\BibitemShut {NoStop}%
\bibitem [{\citenamefont {Kim}, \citenamefont {Payne},\ and\ \citenamefont
  {Zuo}(2012)}]{Kim2012}%
  \BibitemOpen
  \bibfield  {author} {\bibinfo {author} {\bibfnamefont {K.-H.}\ \bibnamefont
  {Kim}}, \bibinfo {author} {\bibfnamefont {D.~A.}\ \bibnamefont {Payne}}, \
  and\ \bibinfo {author} {\bibfnamefont {J.-M.}\ \bibnamefont {Zuo}},\
  }\bibfield  {title} {\enquote {\bibinfo {title} {{Symmetry of piezoelectric
  (1− x)Pb(Mg1/3Nb2/3)O3-xPbTiO3 (x = 0.31) single crystal at different
  length scales in the morphotropic phase boundary region}},}\ }\href {\doibase
  10.1103/PhysRevB.86.184113} {\bibfield  {journal} {\bibinfo  {journal}
  {Physical Review B}\ }\textbf {\bibinfo {volume} {86}},\ \bibinfo {pages}
  {184113} (\bibinfo {year} {2012})}\BibitemShut {NoStop}%
\end{thebibliography}%

\end{document}